\documentclass[conference]{IEEEtran}
\pdfoutput=1

\usepackage[T1]{fontenc}
\usepackage[utf8]{inputenc}

\usepackage{amsmath,amssymb,amsfonts}
\usepackage[export]{adjustbox}
\usepackage{cite}
\usepackage{graphicx}
\usepackage{placeins}
\usepackage{xcolor}

\usepackage{glossaries}
\glsdisablehyper
\newacronym{cir}{\mbox{CIR}}{channel impulse responses}
\newacronym{cots}{\mbox{COTS}}{commercial off-the-shelf}
\newacronym{isac}{\mbox{ISAC}}{integrated sensing and communication}
\newacronym{los}{\mbox{LoS}}{line-of-sight}
\newacronym{sl}{SL}{side link}
\newacronym[plural=\mbox{SDRs},firstplural=software defined \mbox{radios (SDR)}]{sdr}{\mbox{SDR}}{software defined radio}
\newacronym{snr}{\mbox{SNR}}{signal-to-noise ratio}
\newacronym{rx}{Rx}{receiver}
\newacronym{tx}{Tx}{transmitter}
\newacronym[plural=\mbox{UAVs}]{uav}{\mbox{UAV}}{unmanned aerial vehicle}
\newacronym{ue}{UE}{user equipment}
\newacronym{uldl}{UL/DL}{up/down link}

\usepackage{siunitx}
\DeclareSIUnit{\dBm}{\deci\bel m}

\usepackage{hyperref}
\hypersetup{
	pdfauthor={Julia Beuster, Carsten Smeenk, Saw James Myint, Reza Faramarzahangari, Carsten Andrich, Sebastian Giehl, Christian Schneider, Reiner S. Thomä},
	pdftitle={Sounding-Based Evaluation of Multi-Sensor ISAC Networks for Drone Applications: Measurement and Simulation Perspectives},
    pdfkeywords={Integrated sensing and communication (ISAC), 6G, distributed multi-sensor network, unmanned aerial vehicle (UAV), bi-static target reflectivity measurement and simulation, propagation measurements and modeling}
}

\usepackage{orcidlink}

\title{Sounding-Based Evaluation of Multi-Sensor\\ ISAC Networks for Drone Applications:\\ Measurement and Simulation Perspectives}

\IEEEoverridecommandlockouts

\author{
    \IEEEauthorblockN{%
        Julia~Beuster\IEEEauthorrefmark{1}\,\orcidlink{0000-0003-1887-4278},
        Carsten~Smeenk\IEEEauthorrefmark{2}\,\orcidlink{0009-0007-9062-0025},
        Saw~James~Myint\IEEEauthorrefmark{1}\,\orcidlink{0009-0007-3788-7126}, 
        Reza~Faramarzahangari\IEEEauthorrefmark{1}\,\orcidlink{0009-0008-5300-8088},\\
        Carsten~Andrich\IEEEauthorrefmark{1}\,\orcidlink{0000-0002-4795-3517},
        Sebastian~Giehl\IEEEauthorrefmark{1}\,\orcidlink{0009-0008-1672-1351}
        Christian~Schneider\IEEEauthorrefmark{1}\,\orcidlink{0000-0003-1833-4562},
        Reiner~S.~Thomä\IEEEauthorrefmark{1}\,\orcidlink{0000-0002-9254-814X}
    }
    \IEEEauthorblockA{\IEEEauthorrefmark{1}Institute of Information Technology and Thuringian Center of Innovation in Mobility, TU Ilmenau, Germany}
	\IEEEauthorblockA{\IEEEauthorrefmark{2}Fraunhofer Institute for Integrated Circuits IIS, Ilmenau, Germany}
    \thanks{This research has been partially funded by the Federal State of Thuringia, Germany, and the European Social Fund (ESF) under the grants 2017~FGI~0007, 2018~FGR~0082, and 2021~FGI~007, the Federal Ministry of Education and Research of Germany in the projects ``6G-ICAS4Mobility'' (grant number: 16KISK241) and ``KOMSENS-6G'' (grant number: 16KISK125) and by the Free State of Bavaria in the ``DSAI'' project.}
}

\begin{document}

\maketitle

\begin{abstract}
With the upcoming multitude of commercial and public applications envisioned in the mobile 6G radio landscape using \glspl{uav}, \gls{isac} plays a key role to enable the detection and localization of passive objects with radar sensing, while optimizing the utilization of scarce resources.
To explore the potential of future \gls{isac} architectures with \glspl{uav} as mobile nodes in distributed multi-sensor networks, the system's fundamental capability to detect static and dynamic objects that reveal themselves by their bi-static back-scattering needs to be evaluated.
Therefore, this paper addresses simulation- and measurement-based data acquisition methods to gather knowledge about the bi-static reflectivity of single objects including their Micro-Doppler signature for object identification as well as the influence of multi-path propagation in different environments on the localization accuracy and radar tracking performance.
We show exemplary results from simulation models, bi-static reflectivity measurements in laboratory environment and real-flight channel sounding experiments in selected scenarios showcasing the potential of synthetic and measured data sets for development and evaluation of \gls{isac} algorithms.
The presented measurement data sets are publicly available to encourage the academic RF community to validate future algorithms using realistic scenarios alongside simulations models.
\end{abstract}


\section{Introduction}
The increasing accessibility and affordability of \glspl{uav} and their incorporation into mobile radio services (5G and beyond) is paving the way for a variety of future applications in commercial and public services using single \glspl{uav} and \gls{uav} swarms \cite{8918497, 9456851}.
A typical use case is the operation in reconnaissance missions for traffic monitoring or search and rescue particularly in challenging or inaccessible environments such as traffic accidents sites and natural disaster areas.
Furthermore, \glspl{uav} can be used to enhance surveillance measures to protect critical infrastructure, support law enforcement and bolstering public safety.
A key functionality for these application areas is the creation of  comprehensive situation maps of passive dynamic and static objects on the ground and in the air, which facilitates navigation, ensures precise height control especially above challenging terrain and vegetation, enables the implementation of a sense-and-avoid system for collision avoidance, and allows to identify and localize objects of interest.
To ensure this key feature even for visually difficult weather conditions and terrain as well as harsh environments regarding the GNSS reception, supplementary radio sensing is essential.

With regard to the strict weight and space constraints in \glspl{uav} it is not resource efficient to use dedicated hardware for communication-only and sensing-only purposes. 
In contrast, \gls{isac} offers the prospect to exploit resources of already existing mobile radio infrastructure for radar sensing and therefore, lowering the cost and energy consumption \mbox{\cite{9585321, WIP_Kaiserslautern, liu2019joint, 9829746, mandelli2023survey, hexa-x-gap}}.
Two aspects are crucial for the radar performance of distributed multi-sensor \gls{isac} networks \cite{10289611} in identifying, characterizing, and localizing static obstacles and dynamic objects along their motion trajectories:
\begin{enumerate}
    \item the bi-static reflectivity and Micro-Doppler signatures of single objects, and
    \item the impact of multi-path propagation on multi-static radar tracking.
\end{enumerate}
This paper addresses the challenges of acquiring simulation- and measurement-based sounding data sets to evaluate the capabilities of future multi-sensor \gls{isac} networks for \gls{uav} applications.
This includes the presentation of exemplary results from bi-static reflectivity simulations and laboratory measurements in the BiRa facility at TU Ilmenau as well as research using channel modeling and real-world channel sounding in selected \gls{uav} flight scenarios and multi-path environments, which highlight the potential of publicly accessible measurement data sets, as available \mbox{at \cite{EMS_dataset} and \cite{WIP_EMS_BIRA_dataset}}, for development and evaluation of ISAC algorithms.

\section{Multi-sensor \gls{isac} network for \gls{uav} applications}

\begin{figure}[ht]
	\includegraphics[width=\linewidth] {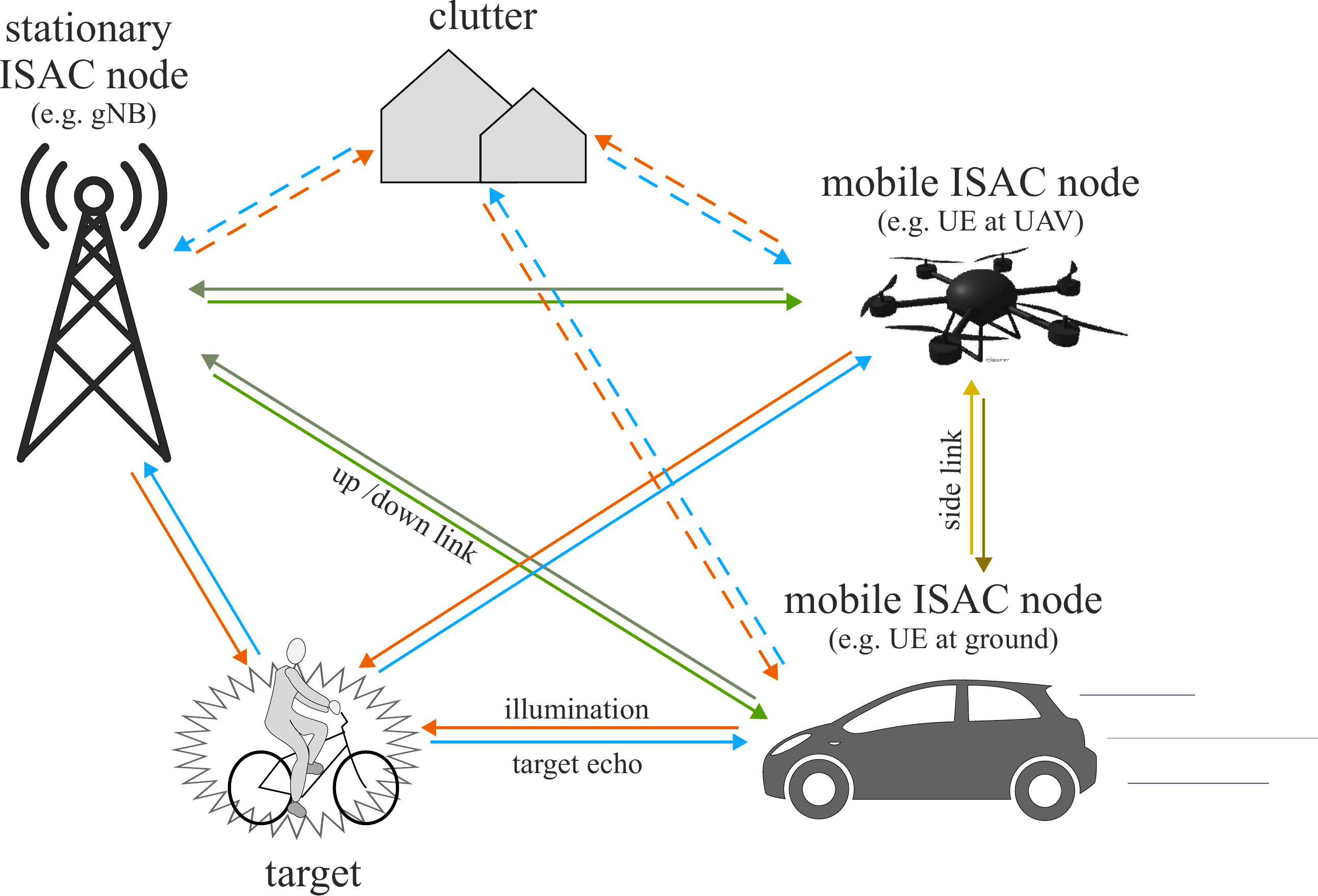}
	\caption{Simplified multi-sensor \gls{isac} network \cite{10289611} with mobile nodes of UE at \glspl{uav} and cars, to illustrate sensing in up/down link scenarios as well as in side link scenarios with direct UE2UE communication.
    }
\label{MS_MIMO_UAV}
\end{figure}

A multi-sensor \gls{isac} network, as introduced in \cite{10289611}, consists of multiple transceiver nodes operating as illuminators and sensors in a distributed infrastructure to detect static obstacles and track position-related parameters of moving objects. 
\mbox{\autoref{MS_MIMO_UAV}} illustrates a basic meshed \gls{isac} system that includes \glspl{uav} as mobile \gls{isac} nodes.
In such a network the \mbox{\gls{ue}} at \glspl{uav} can be used as \mbox{\gls{tx}} for illuminating targets and as \gls{rx} for sensing the echos of targets illuminated by \gls{ue} at \gls{isac} nodes in \mbox{\gls{sl}}, e.g. for mobile nodes autonomously operating in areas without base station coverage, or by base stations in a \gls{uldl} scenario or in a \gls{sl} scenario as auxiliary external illuminator, if the \gls{ue}-based \gls{tx} is too weak.

\begin{figure}[ht]
	\centering
	\includegraphics[width=\linewidth] {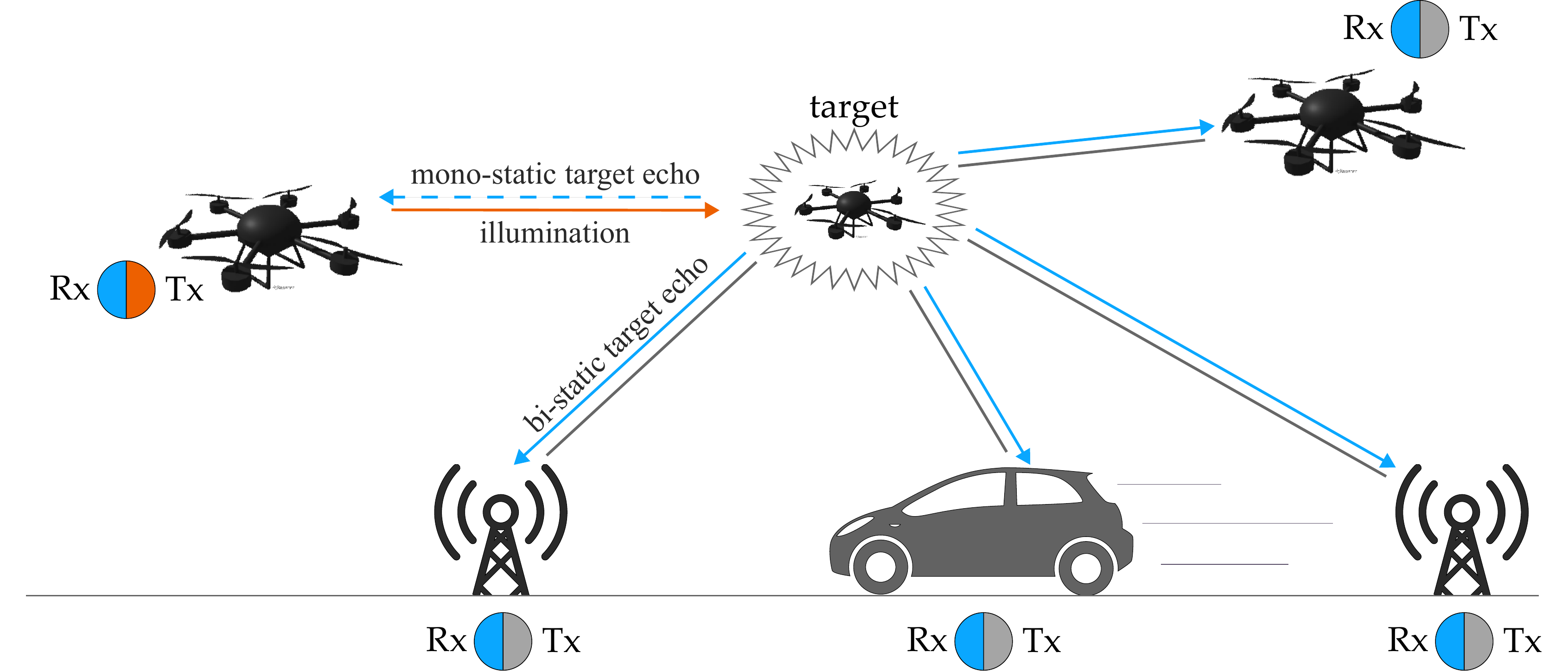}
	\caption{Illustration of multi-sensor architecture with multiple (mobile) distributed \gls{isac} transceiver nodes and a target. One transceiver node operates as illuminator (Tx) with mono-static radar sensing (Rx), while the other nodes operate in Rx-only mode to enable bi-static radar sensing. Typically, all nodes can act as \gls{tx} and \gls{rx}.}
\label{ISAC_meas_setup}
\end{figure}

As shown in \autoref{ISAC_meas_setup}, the widely distributed geometry in a multi-sensor \gls{isac} network allows bi-static radar sensing which provides a substantial increase in target related back-scatter diversity and facilitates the detection of object movement by avoiding Doppler blindness \cite{arxiv_thomae}.
This holds specific importance as \gls{uav} applications necessitate knowledge about dynamic objects, for instance \glspl{uav}, cars, cyclists, pedestrians, and animals, in addition to information about static objects as buildings, bridges, tree canopies, roadways, high-voltage power lines, and wind turbines.

\vspace{1em}
Mobile radio networks rely on an extensive infrastructure to enable high data rates for a high number of connected devices with low latency while considering 
data security and privacy aspects, synchronization and timing mechanisms, link adaption, channel state estimation and multi-user access.
The key for distributed multi-sensor \gls{isac} networks is the exploitation of these mobile radio services on network and radio access level for radar sensing by dual using frequency bands, wave forms, and hardware.
This communication-centric approach entails a variety of challenges for multi-sensor \gls{isac} architectures as resource allocation, multi-sensor link coordination, data security, synchronization and timing \mbox{\cite{8918497, 9456851, 9585321, WIP_Kaiserslautern, liu2019joint, 9829746, mandelli2023survey, hexa-x-gap, 10289611, arxiv_thomae, 9627227}}.
Of special interest for the future use of multi-sensor \gls{isac} networks in \gls{uav} applications are hereby:
\begin{itemize} 
    \item advanced signal processing algorithms to allow the detection of multiple targets and identification even of targets with low reflectivity, e.g. small \glspl{uav}, to characterize their type, size, and shape while suppressing clutter signals in multi-path environments \cite{9896671, 1263228},
    \item real-time capabilities with focus on contiguity in the parameter space to enable the tracking of moving objects using position-related parameter estimation for range, angle, and Doppler \cite{arxiv_thomae,  deoliveira2023bistatic}, 
\end{itemize}
To research the potential of algorithms for their future use in multi-sensor \gls{isac}
networks with mobile \glspl{uav}, knowledge about the single object's reflectivity characteristics is required as well as the consideration of realistic multi-path propagation.

\section{Data Sets for Bi-static Reflectivity and micro-Doppler signatures of Single Objects}

\begin{figure*}[ht]
\centering
\includegraphics[width=\linewidth]{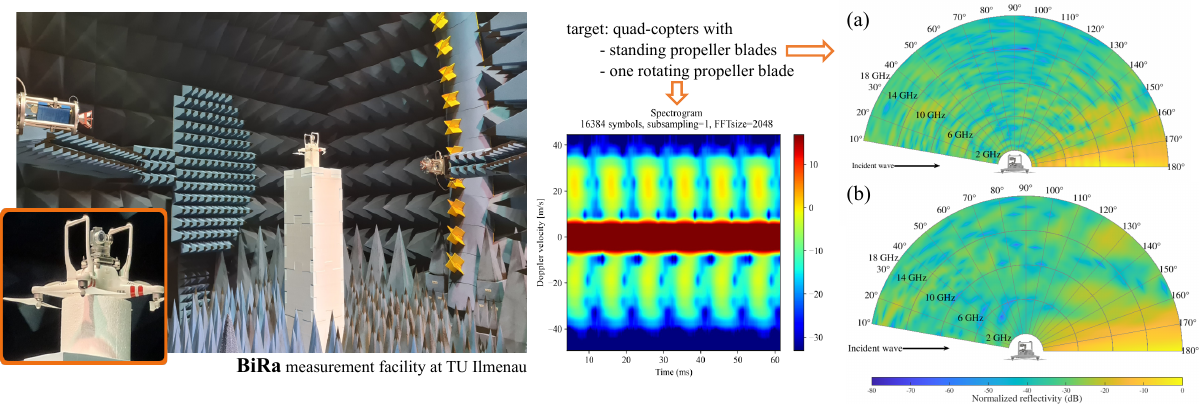}       
\caption{
Impressions of using the BiRa measurement facility at TU Ilmenau to characterize the bi-static reflectivity of multi-copters. 
The illustrated results exemplarily demonstrate the capabilities of BiRa for characterizing the bi-static reflectivity of a small quad-copter, namely a \mbox{DJI Phantom 2}, with propellers at standstill (upper right plot with measured data, lower right plot with simulated data) over the frequency range 2 to 18 GHz for different bi-static angels 10$^{\circ}$ to 180$^{\circ}$ and the Micro-Doppler signature of a quad-copter with one rotating propeller blade (left plot). For more details refer to \cite{arxiv_thomae, WIP_James, WIP_HERALDO}.
}
\label{BIRA}
\end{figure*}

A particularly interesting characteristic of distributed \gls{isac} networks with airborne nodes is the capability to illuminate objects and sense the back-scattered signals from all possible directions in azimuth and elevation allowing the use of bi-static reflectivity for target detection.
Since the reflectivity of an object depends mainly on its geometrical shape, its composed material, and the aspect angles and polarizations of the electromagnetic waves, it provides information about type, size, shape and orientation of objects which is crucial for target identification in real-life scenarios, for instance to separate multiple targets or suppress clutter.
To reduce the complexity of reflectivity as a multidimensional function for target detection, statistical analysis and modeling methods can be applied \cite{James_VRCS, James_Detection}.

Additionally, some targets introduce characteristic micro-Doppler \cite{MicZha2017} by moving object parts as rotating wheels of bicycles, swinging arms of pedestrians, rotating blades of multi-copters, or flapping wings of birds that show in form of time-variant bi-static reflectivity.
Since the structure and the frequency of those local movements differ from one target to another target, those micro-Doppler signatures can be very useful for classifying the detected target.
The knowledge about these objects characteristics can be acquired by simulation and measurement. 

\subsection{Simulation models}
The bi-static reflectivity of objects can be estimated by solving Maxwell’s equations numerically or approximated by calculating the total scattering of decomposed simple shapes.
The commercially available software suits for instance Ansys HFSS, Altair FEKO, and CST Studio Suit, can simulate the reflectivity of given complex object shapes and material compositions and to some extent micro-Doppler with software add-ons. 
Their accuracy mainly depends on the implemented EM solvers, such as method of moments (MoM), finite element method (FEM) and physical optics (PO).
This allows to analyze and characterize a multitude of objects as cars  \cite{James_VRCS}, cyclists \cite{Schwind_bicycle}, and \glspl{uav} \cite{WIP_James, WIP_HERALDO} without the need for cost-intensive measurement campaigns, even though a comparison with measurements is inevitable for validation and optimization of simulation models, as shown \mbox{in \autoref{BIRA}}.

\subsection{Measurements}
Bi-static reflectivity can be measured in a laboratory environment, as introduced in \cite{arxiv_thomae}. 
This bi-static radar test bed (BiRa) in a laboratory facility at TU Ilmenau consists of two pivotable gantries and a turntable to allow illumination and sensing from arbitrary positions within the elevation angles of \SI{0}{\degree} to 115$^{\circ}$ and the azimuth angles of 0$^{\circ}$ to 360$^{\circ}$. 
Therefore, the coverage area is more than a half-hemisphere.
The BiRa test bed facilitates the flexible use of RF measurement systems so that they can be chosen depending on applications, choice of waveforms, bandwidth, and carrier frequencies. 
For example, the current BiRa setup uses a classical VNA for measurements over the frequency range of 2-18 GHz, and a customized 2x2 channel sounder with an instantaneous bandwidth of up to 4 GHz for micro-Doppler measurement in FR1 and FR2~\cite{arxiv_rfsoc}. 

\autoref{BIRA} shows an exemplary setup to measure the bi-static reflectivity of \glspl{uav} in form of multi-copters, which can be used to characterize their static reflectivity with propellers at standstill and identify micro-Doppler signatures introduced by rotating blades. 
These measurements can be used to compare and optimize synthetic data by simulation models with realistic ones.
For more details refer to \cite{arxiv_thomae, WIP_James, WIP_HERALDO}.

Selected sets of the recorded BiRa measurement data are publicly available \mbox{at \cite{WIP_EMS_BIRA_dataset}} to support the use of bi-static reflectivity patterns and micro-Doppler signatures for the detection of vulnerable road users \cite{Schwind_VRU, Schwind_bicycle}, cars \cite{James_VRCS, James_Detection}, \glspl{uav} \cite{arxiv_thomae, WIP_James, WIP_HERALDO}, and other targets of interest in UAV applications.

\section{Test data set acquisition for Multi-Static Radar Sensing in Multi-Path Environments}
\begin{figure}[ht]
\centering
	\includegraphics[width=\linewidth] {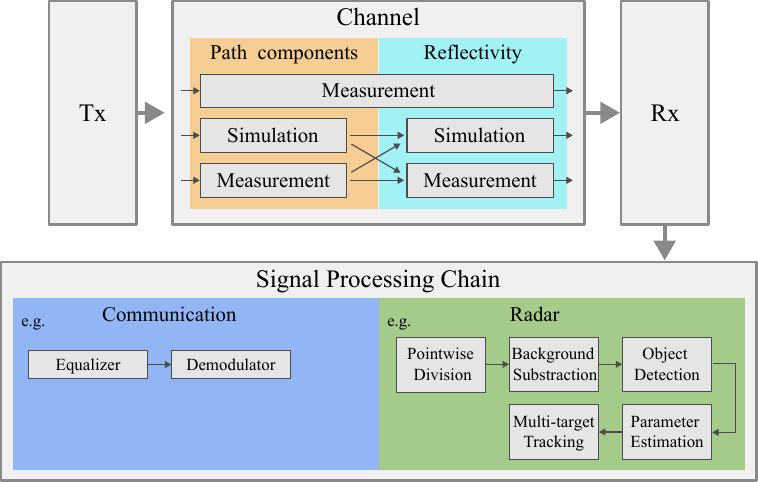}
    \caption{
        Flow diagram of a modular \gls{isac} simulation framework that allows to test algorithms at different stages in the communication and radar signal processing chain. The framework is designed to enable the use of different simulations models and measurement data from real-time channel sounders. For more details refer to \cite{10107507}.  
        }
\label{flowchart}
\end{figure}

Design approaches for basic \gls{isac} system often assume \gls{los} propagation which disregards the influence of propagation in multi-path situations on the visibility of targets, for instance by clutter from obstacles in the surrounding environment, and disables algorithms that proactively exploit multi-path scenarios for radar sensing.
For proper performance evaluation of future \gls{isac} algorithms in the presence of multi-path propagation, it is necessary to emulate the multi-node architecture of distributed \gls{isac} networks in a suitable framework combining exchangeable modules at system and link level.
The former, to emulate the dynamic \gls{isac} network from the variety of nodes and targets shown in \autoref{MS_MIMO_UAV}, the latter to include the radio channel in form of parameters as delay, Direction of Departure (DoD), Direction of Arrival (DoA), and Doppler \cite{arxiv_thomae}.
An extensive, modular framework for testing \gls{isac} signal processing chains and parameter estimation, as introduced in \cite{10107507}, additionally allows to exchange simulation data from channel modeling and measurement data from channel sounding, as shown in \autoref{flowchart}.

\subsection{Simulation}
As multi-sensor \gls{isac} networks estimate the target's state vector using the received signals from multiple distributed sensing nodes, the radio channel needs to change in a spatially and temporally consistent way according to the geometry between the participating transceiver nodes and the target \cite{arxiv_thomae}. 
Therefore, the requirements for channel simulation models differ between radar and communication applications, where multi-path propagation geometry is typically based on statistics as in GSCM.
In order to enable the testing of \gls{isac} algorithms in a controlled, simplified multi-path scenario, it is crucial that the channel modeling is geometrically correct for the links between all nodes, includes knowledge about the target's reflectivity and Micro-Doppler characteristics, and enables clutter modeling. 
The test framework, introduced \mbox{in \cite{10107507}}, is based on such a channel simulation with focus on the continuous streaming of \gls{cir} and contiguity in the parameter space.
This allows the evaluation of future algorithms in a variety of multi-path environments without the need for time-consuming measurements, even though a
validation with measurement data, as shown in \mbox{\autoref{DroneShield_results}}, is needed to assess the real-world performance.

\subsection{Measurement Data}
Measurement test beds that are capable to emulate multi-node \gls{isac} networks for \gls{uav} applications in different environments with static and dynamic obstacles comprise:
\begin{itemize}
    \item a modular system of stationary and mobile transceiver nodes with switchable \gls{rx} and \gls{tx} signal paths (see \mbox{\autoref{ISAC_meas_setup})} and configurable signal parameters, that is flexible in number of nodes, mountable at \glspl{uav} and cars, and allow easy deployment and operation at almost arbitrary positions for ground stations,
    \item highly reliable continuous real-time data recording of \gls{cir} to enable performance evaluation of \gls{isac} algorithms without the need for repetitive measurement campaigns, 
    \item nanosecond-level synchronization accuracy, and 
    \item high accuracy in terms of orientation and positioning for a reliable ground truth of all participating nodes and targets.
\end{itemize}

Initial stages for implementing such a multi-node real-time channel sounder for \gls{isac} scenarios are presented in form of a modular system of distributed synchronized stationary and mobile car-mountable \cite{9128562} and \gls{uav}-mountable transceiver nodes \cite{10133118} for emitter and radar localization.
This test bed uses remote-controlled nodes build from \gls{cots} hardware with MIMO \gls{sdr} transceivers that are configurable in terms of signal parameters and \gls{rx}/\gls{tx} signal paths.
The software architecture licensed under GPLv3  \cite{github_usrp_rxtx} allows highly reliable, continuous real-time data recording and the stationary transceiver nodes with up to 100 MHz instantaneous bandwidth in a frequency range up to 6 GHz are synchronized with nano-second-level accuracy and tailored for easy transport, deployment and low operational effort.
Unfortunately, the introduced light-weight SISO transceiver node for \glspl{uav} (56 MHz instantaneous bandwidth in the frequency range up to 6 GHz with up to 23 dBm EIRP) and the car-mountable MIMO transceiver node (100 MHz instantaneous bandwidth in the frequency range up to 6 GHz with up to 43 dBm EIRP), are only used as \gls{tx} without accurate synchronization. 
Even though the test bed uses centimeter-level positioning as ground truth for the participating nodes and targets, it does not record information about their orientation, which is essential for the antenna positions at mobile \gls{isac} nodes especially at \glspl{uav}.
Nevertheless, the introduced test system showed promising capabilities to emulate a small multi-sensor \gls{isac} network with stationary \gls{isac} nodes, e.g. in a UL/DL scenario, with a flying \gls{uav} as target.

\begin{figure}[ht]
    \centering
        \includegraphics[width=\linewidth]{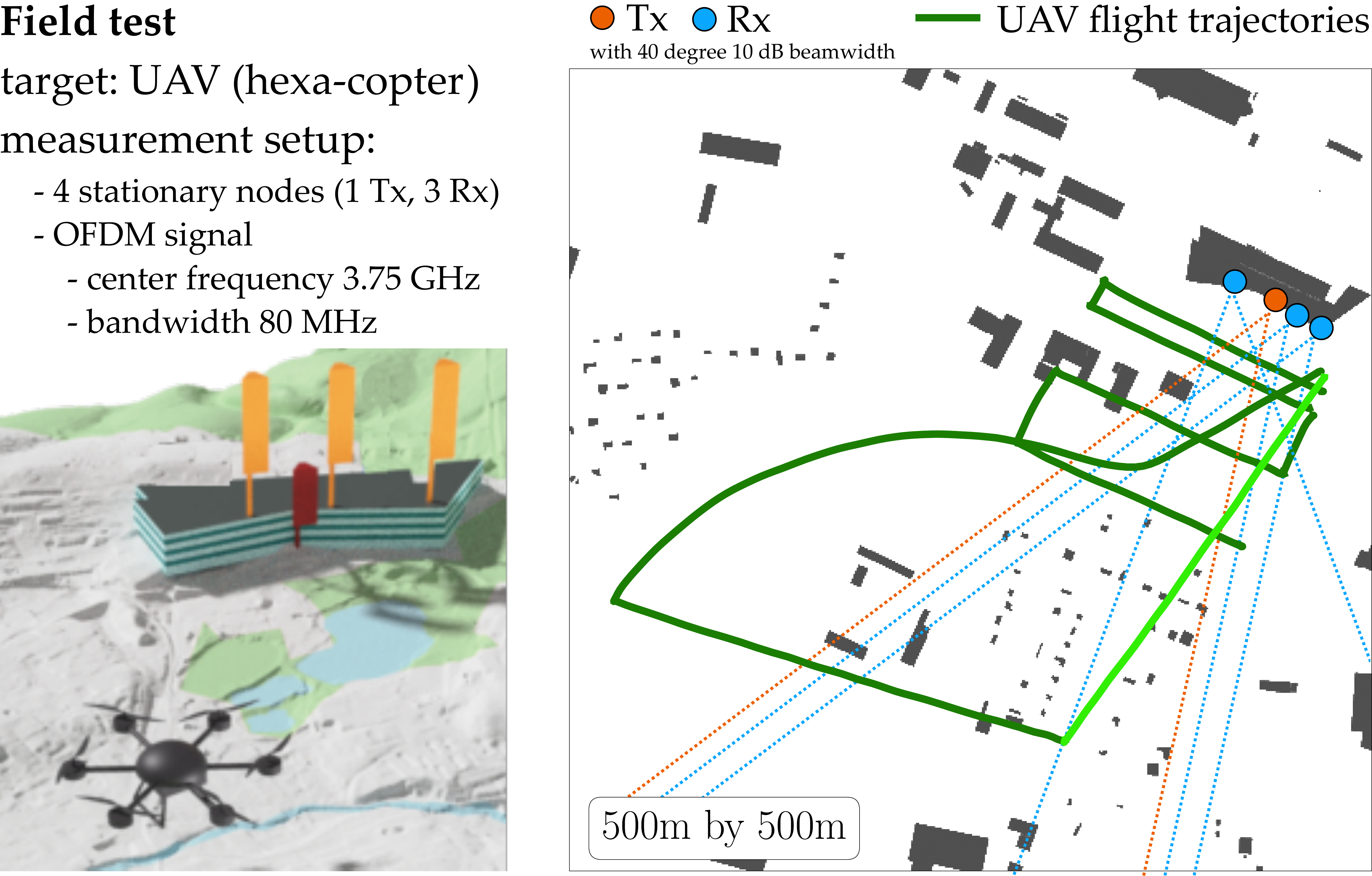}
    \caption{Simplified illustration of field test conducted in \cite{10133118} with 4 stationary transceiver nodes (1 \gls{tx}, 3 \gls{rx}) at the rooftop of a building and a flying \gls{uav} as target. The channel frequency responses for the depicted flight trajectories are publicy available at \cite{EMS_dataset}. Note that the highlighted trajectory \mbox{({\color{green} $\bullet$})} corresponds to the data used in \autoref{DroneShield_results}.}
\label{DroneShield_scenario}
\end{figure}

The exemplary field-test scenario introduced in \cite{10133118} with a publicly accessible data set at \cite{EMS_dataset} comprised a system of \mbox{4 stationary} transceiver nodes with directional antennas at the rooftop of an office building with surrounding infrastructure as neighboring buildings, lamp posts, and parked cars.
The target was a medium-sized hexa-copter flying in parallel and perpendicular trajectories in arrival and departure situations above and below system level, as shown in \mbox{\autoref{DroneShield_scenario}}.
The scenario was tailored to provide conditions and signal parameters similar to those present in \mbox{5G UE} in FR1 and the demonstrated radar range in this scenario was up to \mbox{230 m}.

\begin{figure}[ht]
\centering
    \includegraphics[width=\linewidth]{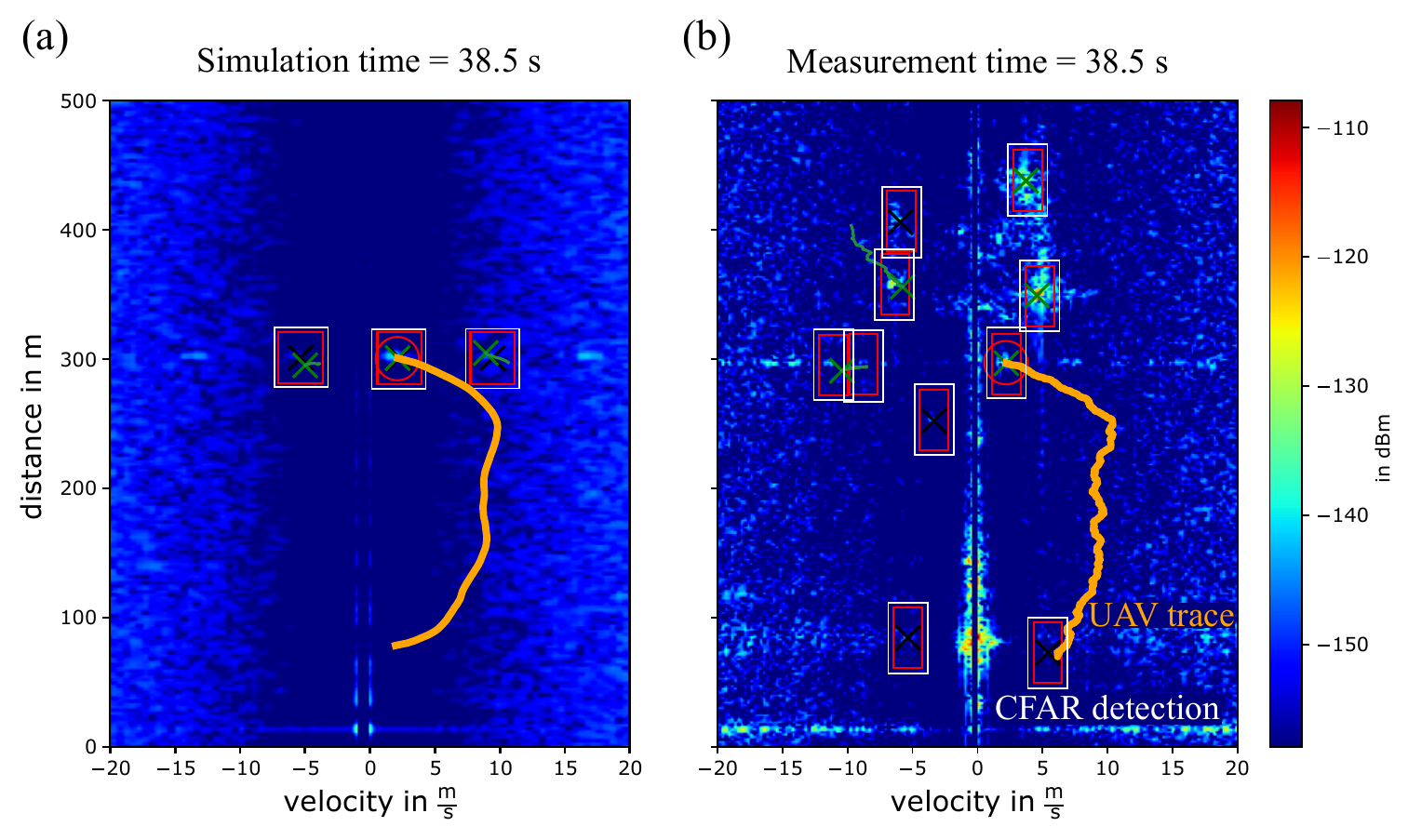}
    \caption{
    Comparison of delay-Doppler-maps of the moving \gls{uav} in a static environment calculated from (a) simulated data \cite{10107507} (b) measured data \cite{EMS_dataset}. The red circle illustrates the ground truth. The orange line is the estimated \gls{uav}'s track. The antenna and \gls{uav} positions are defined by the scenario shown in \autoref{DroneShield_scenario}.}
\label{DroneShield_results}
\end{figure}

\noindent \autoref{DroneShield_results} shows the results from applying exemplary signal processing steps to detect and track the \gls{uav} in one selected flight track of the scenario shown in \autoref{DroneShield_scenario} using the framework and algorithm introduced in \autoref{flowchart} \cite{10107507} with measurement and simulation data.
For the simulated data, the channel considers a multi-path model, antenna and \gls{uav} positions are defined by the measurements, a 3d based reflectivity model with Micro-Doppler signature characterizes the target echo, and multiple uniform distributed cluster components are modeling the static clutter.
Following steps of the radar signal processing chain are applied:
\begin{enumerate}
    \item point-wise division to obtain the radio channel,
    \item background subtraction to remove clutter using exponential background subtraction in the delay-Doppler domain for removal of static signal components and an additional zero-Doppler filter (Notch filter) to suppress highly fluctuating background components,
    \item object detection to classify echo components as targets by applying the threshold detector CA-CFAR, 
    \item parameter estimation to obtain off-grid estimations using fast parabolic interpolation, and
    \item multi-target-tracking (MTT) to stabilize the target detection by compensating short observation losses in predictions and filtering out false alarms, as well as to assign detected echoes.           
\end{enumerate}
For more details refer to \cite{10107507}.
Due to multitude of algorithms, the evaluation and comparison of particular approaches is out of scope in this paper.

To extend the existing measurement setup to meet the requirements of the envisioned multi-node channel sounder, it is necessary to upgrade the inadequately synchronized \gls{tx}-only units to synchronized mobile transceiver nodes for \glspl{uav} and cars and to add an inertial navigation system (INS) to complement the RTK-only ground truth. Furthermore, research regarding the timed schedules for switching of \gls{rx}/\gls{tx} signal paths during one scenario instead of the currently available continuous data recording seems beneficial, for instance for multi-sensor networks at \gls{uav} swarms.

\section{Conclusions}
Within this research we addressed the challenges of distributed multi-sensor \gls{isac} architectures using \glspl{uav} as mobile nodes with focus on the system's key capability to detect objects revealing themselves by their bi-static back-scattering in multi-path environments.
We presented details on the acquisition of data sets using simulation models and measurements to gain insight in the bi-static reflectivity signatures of single objects including their Micro-Doppler and the influence of multi-path propagation scenarios on radar tracking performance.
This included the use of simulation models and measurements in the laboratory facility BiRa at TU Ilmenau as well as the use of a comprehensive \gls{isac} simulation framework and real-flight measurements with a \gls{uav} as target for channel modeling and sounding.

In subsequent work we will use of the introduced, publicly available data sets for development and validation of \gls{isac} algorithms and as training data sets for machine learning .
In addition, we will extend the existing measurement data sets with data from real-world measurement campaigns using a multi-node real-time channel sounder with flying nodes.

\IEEEtriggeratref{19}
\bibliographystyle{IEEEtran}
\bibliography{References}

\end{document}